\noindent\centerline{\bf Comment on Inconsistency of the basic nonadditivity 
of q-nonextensive statistical mechanics, arXiv:0910.3826v1}

\vskip.3cm\centerline{\bf H.J. Haubold}

\vskip.2cm\centerline{Office for Outer Space Affairs, United Nations}
\vskip0cm\centerline{Vienna International Centre, P.O. Box 500} 
\vskip0cm\centerline{A-1400 Vienna, Austria, and} 
\vskip0cm\centerline{Center for Mathematical Sciences}
\vskip0cm\centerline{Pala Campus, Arunapuram P.O., Pala 686 574, India}
\vskip.3cm\centerline{\bf A.M. Mathai} 
\vskip.2cm\centerline{Department of Mathematics and Statistics, McGill
University} \vskip0cm\centerline{Montreal, Canada H3A 2K6, and}
\vskip0cm\centerline{Centre for Mathematical Sciences}
\vskip0cm\centerline{Pala Campus, Arunapuram P.O., Pala 686 574, India}

\vskip.5cm\noindent{\bf Abstract}

\vskip.3cm In a recent paper, Wang. et al. (2009) claim that Tsallis'
nonadditivity of $q$-nonextensive statistical mechanics (Gell-Mann 
and Tsallis 2004, Tsallis 2009) is mathematically
inconsistent and hence one should carefully review
Tsallis' ideas and theory in toto. This present comment is to point out
that the conclusions of Wang et al. (2009) were arrived at by
misinterpreting two basic items and hence the advice of Wang et al.
should be carefully reviewed.

\vskip.5cm\noindent{\bf 1.\hskip.3cm Introduction}

\vskip.3cm Wang et al. (2009) state that Tsallis claims that
``$\alpha$ is a real coupling constant characterizing
nonadditivity''. This is incorrect if ``characterization'' is used
in the mathematical sense. No such statement is implied in Tsallis'
theory, as the present authors understand the subject matter (Gell-Mann 
and Tsallis 2004, Tsallis 2009).
The parameter $\alpha$ characterizes nonadditivity as claimed by Tsallis,
and no claim is made that $\alpha$
is unique. The parameter can be estimated, giving rise to different
values in different contexts, as shown by the many papers on the
topic. It is never claimed to be a universal constant. The parameter
is not given a universal physical interpretation. The present
authors do not believe that nonadditivity 
parameter $\alpha$ can be given a universal physical interpretation
but it is possible to give physical interpretations in terms of
moments which will have different physical meanings in different
contexts.

\vskip.2cm The mathematical inconsistencies pointed out in Wang et
al. (2009) stem from the misreading of statistical independence and
misusing the concept of statistical independence to describe
subsystems of a given system. ``Independence'' is a misnomer and the
events depend on each other through a product probability
property (PPP). Mathai and Haubold (2007) have suggested to replace
the phrase ``statistical independence'' with the statement of
events satisfying ``product probability property (PPP)'' in order to
avoid misuse when the concept is applied to practical situations.
For two events $A$ and $B$, when $P(A\cap B)=P(A)P(B)$, where
$P(\cdot)$ represents the probability of the event $(\cdot)$, the
dependence of $A$ and $B$ is through this product property. When PPP
holds and if logarithm is taken then, naturally,

$$\ln P(A\cap B)=\ln P(A)+\ln P(B).\eqno(a)$$

Thus, a sum is obtained. Shannon's entropy being a logarithmic
function, for any two events satisfying PPP, the entropy in $A\cap
B$ will be the sum of the entropies in $A$ and $B$. In terms of
random variables or probability/density functions, the correct
interpretation is that when the joint probability/density function
is the product of the marginal probability/density functions then
all events on the product space will have the property in (a). In
this case, Shannon's entropy on the joint distribution will be the
sum of the entropies on the individual probability/density
functions.

\vskip.2cm Part of the misinterpretations come from
the use of misleading  notations $+$ for addition by Tsallis and
$\cup$ for union by Wang et al. It is the entropy on the joint
distribution, and if $X(i)$ is used for the entropy on the
probability/density function of the $i$-th variable, then the
entropy on the joint probability/density of $k$ random variables may
be denoted as $X(1,2,...,k)$. If the variables have a general joint
distribution, the entropy of the joint distribution cannot be
expected, and not claimed by anyone, to be the sum of the entropies
on the individual distributions. The claims, and thereby
controversies, are only connected to the situation when the $k$
variables are mutually independently distributed or when PPP holds.
Then only the question of additivity or nonadditivity becomes
meaningful.

\vskip.2cm But Tsallis' entropy and many other $\alpha$-generalized
entropies do not have this additivity property because, basically,
they are not logarithmic functions. For a discussion of such
nonadditive entropies and their mathematical characterizations, see
Mathai and Rathie (1975, 1976).

\vskip.2cm Inconsistencies also come from
misinterpreting subsystem as subsets of a given set, ignoring the
fact that such subsystems need not hold the property (a). Result
corresponding to equation (1) of Wang et al. (2009) will hold for
any number of distributions when the random variables involved are
mutually independently distributed in the sense of joint
probability/density satisfying PPP. Corresponding results are given
in Mathai and Haubold (2007) for the case of more than two
variables, along with mathematical characterizations of Tsallis'
type entropies. Mathematical inconsistencies pointed out in Wang et
al. (2009) are not inconsistencies but misinterpretations of two
items that (1) $\alpha$ is taken as a unique universal constant, (2)
statistical independence of random variables is misinterpreted by
drawing parallel or by comparing to a system and its subsystems,
ignoring PPP. One can have a system and its subsystems satisfying
PPP.

\vskip.2cm Only when $\alpha\rightarrow 1$ Tsallis' entropy and many
other $\alpha$-generalized entropies will have additivity property.
In this sense, physics coming from an $\alpha$-generalized entropy
goes beyond the physics coming from Shannon's entropy or
Boltzmann-Gibbs entropy, giving rise to a wider coverage. If
someone wants to call this wider coverage as ``nonadditivity''
there is no harm in accepting this terminology. Tsallis' statistics
is proved to be a good model to describe many practical situations,
as admitted in Wang et al. (2009), and hence one should go forward
with Tsallis' ideas or any other new idea of ``nonadditivity'' but
at the same time giving proper interpretations which are
mathematically and statistically sound.

\vskip.5cm\centerline{\bf References}

\vskip.5cm\noindent Gell-Mann, M. and Tsallis, C. (eds.): 
{\it Nonextensive Entropy: Interdisciplinary Applications}.
Oxford University Press, New York 2004.

\vskip.5cm\noindent Mathai, A.M. and Haubold, H.J.: Pathway
model, superstatistics, Tsallis statistics and a generalized measure
of entropy. {\it Physica A}, {\bf 375} (2007), 110-122.

\vskip.5cm\noindent Mathai, A.M. and Rathie, P.N.: {\it Basic
Concepts  in Information Theory and Statistics: Axiomatic
Foundations and Applications}, Wiley Halsted, New York 1975.

\vskip.5cm\noindent Mathai, A.M. and Rathi, P.N.: Recent 
Contributions to Axiomatic Definitions of Information and 
Statistical Measures through Functional Equations, in {\it 
Essays in Probability and Statistics}, Eds. S. Ikeda and 
others, Shinko Tsusho, Tokyo 1976, pp. 607-633. 

\vskip.5cm\noindent Tsallis, C.: Nonadditive entropy and 
nonextensive statistical mechanics: An overview after 20 years.
{\it Braz. J. Phys.}, {\bf 39} (2009), 337-356. 

\vskip.5cm\noindent Wang, Q.A., Ou, C.J., El Kaabouchi, A., Chen,
J.C., and Le M\'ehaute, A.: Inconsistency of the basic
nonadditivity of $q$-nonextensive statistical mechanics,
arXiv:0910.3826vl [cond-mat.stat-Mech] 20 Oct.2009.

\bye